\documentstyle[11pt,a4]{article}
\newcommand{\eqalign}[1]{
\null \,\vcenter {\openup \jot \ialign {\strut \hfil $\displaystyle {
##}$&$\displaystyle {{}##}$\hfil \crcr #1\crcr }}\,}
\newcommand{\be}{\begin{equation}}
\newcommand{\ee}{\end{equation}}
\newcommand{\ba}{\begin{array}}
\newcommand{\ea}{\end{array}}
\newcommand{\bea}{\begin{eqnarray}}
\newcommand{\eea}{\end{eqnarray}}
\newcommand{\pa}{\partial}
\newcommand{\cc}{{\rm cos}}
\newcommand{\sS}{{\rm sin}}

\newcommand{\AmS}{{\protect\the\textfont2
  A\kern-.1667em\lower.5
ex\hbox{M}\kern-.125emS}}

\title{ Quantum equivalence of $\sigma$ models related by 
        non Abelian Duality Transformations}

\author{
        L.K. Bal\'azs$^{(1)}$, J. Balog$^{(2)}$, P. Forg\'acs$^{(2)}$,\\ 
N. Mohammedi$^{(2)}$,  
 L.Palla$^{(1)}$\thanks{
        corresponding author E-mail: palla@ludens.elte.hu} 
 and J. Schnittger$^{(2)}$ \\
        {\small $^{(1)}$ Institute for Theoretical Physics,
        Roland E\"otv\"os University,} \\
       {\small  H-1088, Budapest, Puskin u. 5-7, Hungary }\\
       {\small $^{(2)}$ Laboratoire de Math. et Physique Theorique,} \\
       {\small CNRS UPRES-A 6083} \\
       {\small Universit\'e de Tours}\\
       {\small Parc de Grandmont, F-37200 Tours, France}
}

\begin{document}
\maketitle
\begin{abstract}
Coupling constant renormalization is investigated in 2 dimensional $\sigma$ 
models related by non Abelian duality transformations. In this respect it is 
shown that in the one loop order of perturbation theory the duals of
a one parameter family of models, interpolating between the $SU(2)$ principal
model and the $O(3)$ sigma model, exhibit the same behaviour as 
the original models. For the $O(3)$ model also
the two loop equivalence is investigated, and is found to be broken
just like in the already known example  
of the principal model.
\end{abstract}
\vspace{11 mm}
\begin{center}
key words: sigma models, duality, quantum corrections
\end{center}
\vfill\eject
\section{Introduction}
Various target space (\lq T') duality transformations \cite{busch}
connecting two seemingly
different sigma-models
or string-backgrounds are playing an increasingly important role        
nowadays. 
It is assumed that models related by these classical 
transformations are alternative descriptions of the
same physical system (also at the quantum level). The duality transfomations
were originally formulated in the $\sigma$-model description
of the Conformal Field Theory underlying string theory (for a review
see \cite{review}).

Using the $\sigma$-model formulation it has been recently shown that
both the Abelian \cite{alglo2} and the non Abelian
T-duality \cite{curza}
transformation rules can be recovered in an elegant way by
performing a canonical transformation. This clearly shows that models
related by these transformations are {\sl classically} equivalent.
By making  some formal manipulations in the
functional integral without going, however, into the
details of regularization, it is not difficult to argue
that models which are related by duality
transformations correspond to the the same Quantum Field Theory (QFT)
\cite{busch},\cite{frats}. While this may be sufficient for conformal  
invariant string backgrounds \cite{rove} 
(like the gauged \cite{kiri2} or ungauged WZWN \cite{kiri} models) 
when no perturbative quantum corrections are expected, 
 we feel, that from a pure $2$d field theory point of view,  
the question of quantum equivalence between  sigma
models  related by duality transformations deserves further study. 

Concentrating mainly on Abelian duality,   
such a study was initiated in ref.\cite{bfhp}, where   
 the various sigma models were treated as "ordinary" 
(i.e. not necessarily conformally 
invariant) two dimensional quantum field theories in the framework of 
perturbation theory. Working in a field theoretic rather than string
theoretic framework i.e. working without the dilaton on 
a flat, non dynamical $2$ space,  it was shown on a 
number of examples that the `naive' (tree level) T-duality
transformations in 2d $\sigma$-models cannot be
exact symmetries of the quantum theory. The `naive' Abelian duality
transformations 
are correct to one loop in perturbation theory
\cite{bfhp},\cite{haag}, they break down in general,
however, at the two loop order, and to promote them to full quantum symmetries 
some non trivial modifications are needed \cite{bfhp2}, \cite{kalo}. 
 These conclusions were reached by analyzing 
and comparing various $\beta$ functions in the original and dual theories.

The aim of this paper is to repeat as much of this program as possible for 
sigma models related by non Abelian duality transformations \cite{freeto}, 
\cite{osqu},\cite{alglo},\cite{givro}. 
Non Abelian 
duality is a special case of the so called Poisson-Lie T duality 
\cite{klim},\cite{klim2}, 
which 
generalizes the concept of T duality for sigma models without isometries. The
motivation for this investigation came from several directions. First of all 
quantum equivalence among sigma models related by non Abelian duality has 
some problems even in the conformal invariant case, as there are examples 
\cite{vene} where non Abelian duality is mapping a conformal invariant 
background to a non conformal dual. The second, \lq non conformal' motivation 
is the discovery 
\cite{bfhp},\cite{subtu} 
that the relation 
between the $SU(2)$ principal model and its non Abelian dual shows the same 
features as in the case of Abelian duality: in the one loop order the two 
models are equivalent while at two loops the dual is not renormalizable in the
usual, field theoretic sense. The investigation of this problem is also 
made urgent by one of the results of \cite{tim}. In ref.\cite{tim} exact 
$S$ matrices were proposed for a particular class of $2d$ models and in an 
appropriate limit these $S$ matrices yield a non perturbative $S$ matrix, 
which can be associated to   
the non Abelian dual of the $SU(2)$ principal model, and  
which is {\sl identical}
to the well known $S$ matrix of the principal model.  

Therefore, in this paper, we consider a one parameter family of sigma models 
interpolating between the $SU(2)$ principal model and the $O(3)$ sigma 
model together with the non Abelian dual of this family, and investigate 
their renormalization. The interest in this one parameter family comes from 
two sources: on the one hand it provides a convenient laboratory to compare 
in a more general setting 
the renormalization of sigma models connected by non Abelian duality, and on 
the other, by enlarging the parameter space of the principal model it may 
provide a sufficient generalization where the two loop renormalizability of 
the dual model is restored. (This phenomenon was recently shown to happen in
some $SL(3)$ sigma models and their Abelian duals \cite{hkp}). 
As we show in
detail, for the general member of this family, the renormalization 
of the duals in the one loop order leads to the same $\beta $ functions as in
the case of the original models. However for the only  case  
besides the $SU(2)$ principal model, when the complexity of the 
two loop analyzis becomes tractable, 
namely for the $O(3)$ model,  we find that the dual is not 
renormalizable in this order. 

The paper is organized as follows: in sect.2 we describe in some detail the 
two sets of models which are related to each other by non Abelian duality 
transformation. In sect.3 we discuss some aspects of 
the canonical transformation 
implementing this duality transformation in the classical theory. In sect.4 
we give a short summary of the renormalization procedure used and apply it 
in detail to our models. We discuss the results 
and make our conclusions in 
sect.5. The somewhat complicated expressions for the components 
of the generalized Ricci tensor, that determine the one loop counterterm, are 
collected in the Appendix.

\section{The dually related models}
We choose the \lq\lq original" model from the class of \lq deformed 
principal' models, the Lagrangian of which can be written as:
\be\label{osm}
{\cal L}=\frac{1}{2}R_{ab}L^a_{\ i}L^b_{\ j}\pa_\mu\xi^i\pa^\mu\xi^j,
\ee
where $\xi^j$, $j=1,...{\rm dim}{\bf G}$, parametrize the elements, 
$G$, of a group, ${\bf G}$,
 and 
$L^a_{\ i}$ denote the components of the left invariant Maurer Cartan one 
form:
\be\label{ldef}
L^a_{\ i}=\frac{1}{\omega}{\rm tr}
\bigl(\lambda^a G^{-1}\frac{\pa G}{\pa\xi^i}\bigr).
\ee
($\lambda^a$ stand for the generators of the Lie algebra of ${\bf G}$: 
$[\lambda^a,\lambda^b]=f^{abc}\lambda^c$, normalized according to 
${\rm tr}(\lambda^a\lambda^b)=\omega\delta^{ab}$. The inverse of $L^a_{\ i}$ 
is denoted by $L^i_{\ a}$, and we frequently use the abbreviation $L^a_{\mu}=  
L^a_{\ i}\pa_\mu\xi^i$). In Eq.~(\ref{osm}) $R_{ab}$ is a constant (i.e.
$\xi^i$ independent), symmetric $R_{ab}=R_{ba}$ matrix, that describes the
deviation of the model from the 
${\bf G}_L\times{\bf G}_R$ invariant principal model, which 
 is obtained if $R_{ab}\sim \delta_{ab}$. 
In the general case, the presence of $R_{ab}$ breaks the symmetry of the 
principal model to ${\bf G}_L\times {\bf H}_R$, where the actual form of the 
subgroup ${\bf H}_R$ depends on the actual form of $R_{ab}$. $R_{ab}$ also
appears in the equations of motion following from Eq.~(\ref{osm}):
\be\label{oem}
R_{ac}(\pa_\mu L^{a\mu})-R_{ab}L^a_\mu L^{k\mu}f^{bkc}=0.
\ee
Therefore, although even in the general case $L^a_\mu$ form 
curvature free currents 
on account of the definition, Eq.~(\ref{ldef}), 
\be\label{ocf}
\pa_\mu L^a_\nu-\pa_\nu L^a_\mu +f^{abc}L^b_\mu L^c_\nu =0,
\ee
they are no longer conserved. Note, however, that the Noether currents:
\be\label{noeth}
C^a_\mu=R_{cd}N^c_{\ a}L^d_\mu ,\qquad 
N^c_{\ a}=\frac{1}{\omega}{\rm tr}(\lambda^cG^{-1}\lambda_aG),
\ee
are conserved as a result of ${\cal L}$ being invariant under the left 
${\bf G}$ transformations: $G\rightarrow WG$, $W\in {\bf G}$.

The non Abelian dual of the model in 
Eq.~(\ref{osm}) can be written as \cite{givro},\cite{klim},\cite{klim2},
\cite{sfet},\cite{ung}:
\be\label{dsm}
{\cal L}^d=\frac{1}{2}\pa_+\chi^a[M^{-1}]^{ab}\pa_-\chi^b,
\ee
where $\chi^a$ are coordinates on the Lie algebra of ${\bf G}$, 
$\pa_\pm\chi^a=\pa_\tau\chi^a\pm\pa_\sigma\chi^a$, and the matrix $M$ is 
given as:
\be\label{mdef}
M^{ab}=R^{ab}+f^{abc}\chi^c.
\ee
The transformation connecting ${\cal L}$ and ${\cal L}^d$ is a special case
of the Poisson Lie T duality \cite{klim},\cite{klim2}, 
when the dual group is Abelian. 

The actual pair of dual models where we investigate the question of their
quantum equivalence is obtained from Eqs.~(\ref{osm}-\ref{mdef})
by choosing ${\bf G}=SU(2)$ and 
\be\label{Rexp}
R_{ab}=\frac{1}{\lambda}{\rm diag}(1,1,1+g),
\ee 
where $\lambda$ is the coupling constant (which is there even for the principal
model) and $g$ is the deformation parameter. 
Using the Euler angles ($\phi$,$\theta$,$\psi$) to pa\-ra\-met\-ri\-ze
the elements of $SU(2)$, $G$ is written as
\be\label{euler}
G=e^{\phi\tau^3}e^{\theta\tau^1} e^{\psi\tau^3}\, ,
\ee
(where $\tau^a=\sigma^a/(2i)$, with $\sigma^a$ being the standard Pauli 
matrices), and one readily obtains 
\be\eqalign{
\label{lexp}
L^3_\mu &=\pa_\mu\psi +\pa_\mu\phi\cos\theta ,\cr 
L^1_\mu &=\pa_\mu\theta\cos\psi +\pa_\mu\phi\sin\theta\sin\psi ,\cr 
L^2_\mu &=-\pa_\mu\theta\sin\psi +\pa_\mu\phi\sin\theta\cos\psi .}
\ee
Then the Lagrangian of the deformed $SU(2)$  model (\ref{osm})  becomes
\be\label{lagrsig2}
\eqalign{
{\cal L}=&{1\over2\lambda}\bigl\{ (\pa_\mu\theta)^2+(\pa_\mu\phi)^2\bigl( 1+g
\cos^2\theta\bigr)+\cr
&(1+g)(\pa_\mu\psi)^2+2(1+g)\pa_\mu\phi\pa^\mu\psi\cos\theta\bigr\}\,.
}\ee
Note that for $g=-1$ the $\psi$ field decouples and Eq.~(\ref{lagrsig2}) 
reduces to the Lagrangian of the $O(3)$ sigma model. Thus Eq.~(\ref{lagrsig2}) 
describes a one parameter family of models interpolating between the $SU(2)$ 
principal model ($g=0$) and the $O(3)$ one. Using the explicit form of 
$R_{ab}$ in Eq.~(\ref{oem}) shows that
\be\label{su2em}
(1+g)\pa_\mu L^{3\mu }=0,\qquad 
\pa_\mu L^{{1\atop 2}\mu }\pm gL^3_\mu L^{{2\atop 1}\mu}=0;
\ee
i.e. (for $g\ne -1$) $L^3_\mu $ is conserved. This conservation is the 
manifestation of Eq.~(\ref{lagrsig2}) being invariant under the 
$\psi\rightarrow\psi +\psi_0$ translation. Since this translation acts on 
$G$, Eq.~(\ref{euler}), on the right, the total symmetry group 
of the deformed $SU(2)$ model is $SU(2)_L\times U(1)_R$.  

As only $R_{ab}$ is singular for
$g=-1$, but $M^{ab}$ is not, the dual model is readily defined for all 
$g\ge -1$. Rescaling the $\chi^a$ fields appropriately and introducing
the variables:
\be\label{roald}
\chi^1=\rho\cos\alpha ,\quad \chi^2=\rho\sin\alpha ,\quad
\chi^3=z,
\ee
the Lagrangian of the dual model assumes the form:
\be\eqalign{
\label{dualag}
{\cal L}^d=\frac{1}{2\lambda D}&\Bigl( (1+g+\rho^2)(\pa_\mu\rho)^2+
(1+g)\rho^2(\pa_\mu\alpha)^2+(1+z^2)(\pa_\mu z)^2+\cr
&2z\rho\pa_\mu z\pa^\mu\rho 
-2\epsilon_{\mu\nu}\bigl( (1+g)z\rho\pa^\mu\rho\pa^\nu\alpha +
\rho^2\pa^\mu\alpha\pa^\nu z\bigr)\Bigr)\, ,\cr
&D=(1+g)(1+z^2)+\rho^2\, .}
\ee  
Note that for $g=0$ the substitution $\rho=r\sin\gamma$, $z=r\cos\gamma$,  
really converts Eq.~(\ref{dualag}) into the well known non Abelian dual of
the $SU(2)$ principal model, while for $g=-1$, when (after discarding a 
total derivative) the $\alpha$ field decouples, it reduces to a 
purely metric model:
\be\label{o3nad}
{\cal L}^d_{O(3)}=\frac{1}{2\lambda }\Bigl( (\pa_\mu\rho)^2+
\frac{1+z^2}{\rho^2}(\pa_\mu z)^2+
2\frac{z}{\rho}\pa_\mu z\pa^\mu\rho\Bigr)\, ,
\ee
which may be called the non Abelian dual of the $O(3)$ sigma model. (This 
particular non Abelian dual is a special case of the coset 
examples discussed in \cite{klim3}). The only 
manifest symmetry of the dual Lagrangian, Eq.~(\ref{dualag}), is $U(1)$, 
corresponding to the $\alpha\rightarrow\alpha +\alpha_0$ translation.

\section{The canonical transformation}
Before investigating the equivalence of the quantized versions of 
Eqs.~(\ref{lagrsig2}) and (\ref{dualag}) we review the canonical transformation
that connects these models classically. In doing so we clarify the relation
between the symmetries of the dually related models and also make a minor 
observation on the interpretation of the transformation itself.

The canonical transformation implementing non Abelian duality in case of the
principal models was described in \cite{curza},\cite{oal}, 
while for the \lq left invariant 
models' (which Eq.~(\ref{osm}) belongs to) it was worked out in \cite{sfet}. 
Poisson-Lie T duality is by definition a canonical transformation and an 
expression for the generating functional is given in \cite{klim2}. 

We define the canonical momenta of the original and dual models in the
usual way:
\be\label{mom}
p_j=\frac{\pa{\cal L}}{\pa\dot{\xi}^j}=R_{ab}L^a_{\ j}L^b_\tau ,\quad 
\pi_a=\frac{\pa{\cal L}^d}{\pa\dot{\chi}^a}=\frac{1}{2}\bigl( 
[M^{-1}]^{ab}\pa_-\chi^b+[M^{-1}]^{ba}\pa_+\chi^b\bigr)\, .
\ee
Then the canonical transformation $(\xi^i,p_j)\rightarrow (\chi^a,\pi_b)$ 
following from the generating functional:
\be\label{genfu}
F[\chi^a,\xi^i]=\oint\limits_{S^1}\chi^aL^a_{\ i}\pa_\sigma\xi^i\, ,
\ee
can be written as:
\be\label{cf1}
\pi_a=\frac{\delta F}{\delta \chi^a}=L^a_{\ i}\pa_\sigma\xi^i\, ,
\ee
\be\label{cf2}
p_j=-\frac{\delta F}{\delta \xi^j}=L^a_{\ j}\pa_\sigma\chi^a
-f^{abc}\chi^aL^b_{\ i}L^c_{\ j}\pa_\sigma\xi^i\, .
\ee
(The Maurer Cartan condition, Eq.~(\ref{ocf}), is used 
in writing Eq.~(\ref{cf2})). Note -- as was pointed out in \cite{sfet} 
-- that both 
the generating functional and the canonical transfomations have the same form
as in the case of the principal model. However, as we show below, in contrast
to the principal model, in the general case, when $R_{ab}\not\sim\delta_{ab}$, 
they map two curvature free (but not necessarily conserved) currents -- 
that are at the starting point of Poisson Lie duality \cite{klim}
-- into each other. 
In the original model this current is of course $L^a_\mu$. In the dual model,
we define the \lq dual current' by
\be\label{ducur}
J^a_-=[M^{-1}]^{ad}\pa_-\chi^d,\qquad J^a_+=-\pa_+\chi^d[M^{-1}]^{da},
\ee after  
observing that using the simple identity: 
\be \frac{\pa [M^{-1}]^{ab}}{\pa\chi^c}=
-[M^{-1}]^{ad}f^{dec}[M^{-1}]^{eb},\ee  the equations of motion following from 
Eq.~(\ref{dsm}) can be interpreted as the curvature free conditions for 
$J^a_\mu $. 

Using the definition of $\pi_a$, Eq.~(\ref{mom}), and 
Eq.~(\ref{ducur}) we get $\pi_a=-J^a_\sigma $, thus the first of the canonical
transformations, Eq.(\ref{cf1}), indeed identifies (up to a sign) the spatial
components of $L^a_\mu $ and $J^a_\mu $. Multiplying both sides of 
Eq.~(\ref{cf2}) by $L^i_{\ c}$ and exploiting Eq.~(\ref{cf1}) leads to 
\be\label{cf22}
L^i_{\ c}p_i=\pa_\sigma\chi^c -f^{abc}\chi^a\pi_b\, .
\ee
It follows from the definition of $p_i$, Eq.~(\ref{mom}), that 
the left hand side of this equation is nothing but $R_{bc}L^b_\tau $. A 
simple computation, using the expression of $\pi_c$ in terms of $J^a_\pm$, 
as well as the obvious identities 
\be
R_{ab}=\frac{1}{2}(M^{ab}+M^{ba});\qquad
f^{abc}\chi^c=\frac{1}{2}(M^{ab}-M^{ba}); 
\ee
which follow from the definition, Eq.~(\ref{mdef}), confirms, that the right 
hand side of Eq.~(\ref{cf22}) can indeed be written as $-R_{ac}J^a_\tau $. 
Thus the canonical transformation really connects the curvature free (but 
in general non conserved) $L^a_\mu$ and $J^a_\mu$:
\be\label{curmap}
L^a_\sigma =-J^a_\sigma ,\qquad R_{bc}L^b_\tau =-R_{bc}J^b_\tau\, .
\ee
Note also that $R_{ab}J^b_\mu$ is related in an interesting way to the 
topological current $T^a_\mu =\epsilon_{\mu\nu}\pa^\nu\chi^a$ and the \lq 
would be' Noether current, $V^a_\pm$,
\be
V^a_+=\frac{1}{2}\pa_+\chi^b[M^{-1}]^{bc}f^{cda}\chi^d,\quad 
V^a_-=\frac{1}{2}f^{cda}\chi^d[M^{-1}]^{cb}\pa_-\chi^b,
\ee
that corresponds to the $\chi^a\rightarrow\chi^a+f^{abc}\chi^b\omega^c$ 
transformation:
\be
-R_{ab}J^b_\pm =2V^a_\pm+T^a_\pm\, .
\ee

Eq.~(\ref{cf22}) also imply that the special combinations of the original 
variables and momenta, $q_c\equiv L^i_{\ c}p_i$, (which are nothing but 
$R_{bc}L^b_\tau $), become local in the dual model. Using the basic Poisson 
brackets $\{\xi^i(\sigma ),p_j(\sigma^\prime )\} =\delta_{ij}
\delta (\sigma -\sigma^\prime )$ it is straightforward to show that the 
bracket among the $q_a$-s in the original model has the form 
$\{ q_a(\sigma ),q_b(\sigma^\prime )\} =f^{abc}q_c(\sigma )
\delta (\sigma -\sigma^\prime )$. As the transformation between the original 
and dual models is canonical, we get the same, if we use 
$\{\chi^a(\sigma ),\pi_b(\sigma^\prime )\} =\delta_{ab}
\delta (\sigma -\sigma^\prime )$  and identify $q_c$ with the quantities on 
the right hand side of Eq.~(\ref{cf22}). However the integrals $\int d\sigma
(\pa_\sigma\chi^c -f^{abc}\chi^a\pi_b)$ generate local, conserved charges in 
the dual model only for those special values of $c$, (if there is any), for 
which $R_{cb}L^b_\mu$ is conserved (see Eq.~(\ref{oem})).    
   
In the deformed $SU(2)$ model only $L^3_\mu$ is conserved; and since 
$L^i_{\ 3}p_i=p_\psi$, the conserved $U(1)_R$ charge is $\int d\sigma p_\psi $.
In the dual model, for $c=3$, in terms of the new variables, Eq.(\ref{roald}), 
the right hand side of Eq.~(\ref{cf22}) can be written as 
$\pa_\sigma z-\pi_\alpha $. Thus (discarding the uninteresting integral of 
a total derivative) we see that the \lq image' of the $U(1)_R$ charge in the 
dual model is the conserved \lq $\alpha $ charge' $\int d\sigma\pi_\alpha $.
 This explaines the 
$\alpha\rightarrow\alpha +\alpha_0 $ symmetry of Eq.~(\ref{dualag}). The 
charges of the left Noether currents, $\int C^a_\tau d\sigma $, of the 
original model become non local in the dual, since $C^a_\tau$ can be written 
as $C^a_\tau =N^c_{\ a}(\xi )L^i_{\ c}p_i=N^c_{\ a}(\xi )  
(\pa_\sigma\chi^c -f^{dbc}\chi^d\pi_b)$. 
 
It is interesting to understand what happens for 
$g=-1$, i.e. to study in more details the
canonical transformation connecting the $O(3)$ sigma model and its non 
Abelian dual. The problem with this is that the 
generating functional, $F[\xi^i,\chi^a]$, Eq.~(\ref{genfu}), is independent 
of $g$, yet for $g=-1$ the $\psi$ field of the $O(3)$ and the $\alpha $ field
of the dual models decouple. In the Hamiltonian formalism, 
these decouplings can be handled in general  
by the procedure described in \cite{klim3}. 
In the present case,
the canonical transformation connecting these two models can be described 
by the following generating functional
\be\label{o3gen}
F=\oint\, d\sigma\,\{z\psi^\prime+z\phi^\prime\cc\theta+
\rho\theta^\prime\cc\psi+\rho\phi^\prime\sS\theta\sS\psi\}\,,
\ee
which  is obtained formally from Eq.~(\ref{genfu}) by setting $\alpha =0$. 
However $\psi$ is not an independent field now, but is rather 
a functional of the 
other fields (and their derivatives), as determined from
\be\label{o3egy}
{\delta F\over\delta\psi}=0.
\ee
Thus $\psi$ satisfies 
\be\label{o3ket}
\rho\phi^\prime\sS\theta\cc\psi-\rho\theta^\prime\sS\psi=z^\prime 
\ee
and could be expressed algebraically in terms of the other fields. However 
we do not need this explicit expression, 
as Eq.~(\ref{o3egy}) guarantees that for the canonical 
transformation of the other fields, $\psi$ appears as an independent field: 
\be\label{o3har}
\eqalign{
\pi_z={\delta F\over\delta z}&=\psi^\prime+\phi^\prime\cc\theta\,,\cr
\pi_\rho={\delta F\over\delta\rho}&=\theta^\prime\cc\psi+
\phi^\prime\sS\theta\sS\psi\,,\cr
p_\phi=-{\delta F\over\delta\phi}&=(z\cc\theta)^\prime+
(\rho\sS\theta\sS\psi)^\prime\,,\cr
p_\theta=-{\delta F\over\delta\theta}&=z\phi^\prime\sS\theta+
(\rho\cc\psi)^\prime-\rho\phi^\prime\cc\theta\sS\psi\,.\cr}  
\ee
>From these equations, using Eq.~(\ref{o3ket}), one readily obtains 
the derivatives of the original coordinates:
\be\label{o3negy}
\eqalign{
\theta^\prime&=-{z^\prime\over\rho}\sS\psi+\pi_\rho\cc\psi\,,\cr
\sS\theta\phi^\prime&={z^\prime\over\rho}\cc\psi+\pi_\rho\sS\psi\,,\cr}
\ee
and the original momenta:
\be\label{o3ot}
\eqalign{
p_\phi=\sS\theta&[\cc\psi(\rho \pi_z-z\pi_\rho)+\sS\psi(\rho^\prime+{zz^\prime
\over\rho})]\,,\cr
p_\theta=&[-\sS\psi(\rho \pi_z-z\pi_\rho)+\cc\psi(\rho^\prime+{zz^\prime
\over\rho})]\,.\cr}
\ee
Eqs.~(\ref{o3negy},\ref{o3ot}) give  
the required 
transformation between the $O(3)$ sigma model and its non Abelian dual.
 
Finally we discuss briefly the question of potential quantum corrections 
to the canonical transformations. Since the generating
functional, Eq.~(\ref{genfu}), is non linear in the variables of the original 
model, one may expect that it receives quantum corrections \cite{ghand},  
when implemented in the functional integral. 
In quantum theory, using Schr\"odinger wave functional techniques, a
{\sl formal} equivalence between the two theories can be established if the
energy-momentum eigenfunctionals of the dual theory, $\Psi_{E,p} [\chi^a]$, 
and those of the original theory, $\Phi_{E,p} [\xi^i]$, are related to each 
other by a non linear functional Fourier transformation \cite{ghand},
\cite{curza}:
\be\label{qucan}
\Psi_{E,p} [\chi^a]=N(E,p)\int\prod\limits_{i=1}^{{\rm dim}{\bf G}}
{\cal D}\xi^i{\rm e}^{i\tilde{F}[\xi^i ,\chi^a ]}\Phi_{E,p} [\xi^i].  
\ee
If $\tilde{F}[\xi^i ,\chi^a ]$ coincides with the classical generating 
functional, $F[\xi^i ,\chi^a ]$, then the classical transformation receives 
no quantum corrections. Let us make the ansatz that indeed 
Eq.~(\ref{qucan}) is 
valid with the classical generating functional, and then verify that, 
as required, an eigenfunction $\Phi_{E,p} [\xi^i]$ of the original
Hamiltonian becomes transformed into an eigenfunction $\Psi_{E,p} 
[\chi^a]$ of the dual one. 
As a result of  
Eq.~(\ref{cf22},\ref{curmap}), the exponentiated classical 
generating functional 
satisfies the functional differential equations:
\be\label{qcurmap}
\eqalign{
&J^a_\sigma \Bigl[ \pi^b\equiv -i\frac{\delta}{\delta\chi^b}\Bigr]
{\rm e}^{iF[\xi^i ,\chi^a ]} 
=-L^a_\sigma [\xi^i]{\rm e}^{iF[\xi^i ,\chi^a ]},\cr
&R_{bc}J^b_\tau \Bigl[ \chi^a,\pi^b\equiv -i\frac{\delta}{\delta\chi^b}\Bigr]
{\rm e}^{iF[\xi^i ,\chi^a ]}
=L^j_{\ c} \bigl[\xi^i\bigr] i\frac{\delta}{\delta\xi^j}
{\rm e}^{iF[\xi^i ,\chi^a ]}
\equiv -R_{bc}L^b_\tau 
{\rm e}^{iF[\xi^i ,\chi^a ]}
\, ,}
\ee
and as a consequence, $\Psi_{E,p}$ and $\Phi_{E,p}$ obey e.g.
\be\label{qcurmap2}
J^a_\sigma\Psi_{E,p} [\chi^a]=-N(E,p)\int\prod\limits_{i=1}^{{\rm dim}
{\bf G}}
{\cal D}\xi^iL^a_\sigma
{\rm e}^{iF[\xi^i ,\chi^a ]}\Phi_{E,p} [\xi^i],  
\ee 
(plus a similar equation where the space components of the currents are 
replaced by $R\ \otimes $ the corresponding time components). Now, 
at least for the
deformed $SU(2)$ model, Eq.~(\ref{lagrsig2}), the Hamiltonian is a simple 
quadratic expression of the $L_\mu^a$-s (which, in turn, are expressed in 
terms of $p_j$ and $\xi^i$):
\be
H=\frac{1}{2\lambda}\int d\sigma \Bigl[\sum\limits_{a=1}^3\bigl( 
(L_\tau^a)^2+(L_\sigma^a)^2\bigr) +g\bigl(  
(L_\tau^3)^2+(L_\sigma^3)^2\bigr)\Bigr]\,.
\ee
(The Hamiltonian, corresponding to the dual model, Eq.~(\ref{dualag}), 
is obtained by using Eq.~(\ref{curmap})). 
Therefore, applying twice the argument leading to Eq.~(\ref{qcurmap2}),
we easily verify that Eq.~(\ref{qucan}) with $\tilde
F[\xi^i ,\chi^a]=F[\xi^i ,\chi^a]$ has the property we wanted to 
check, just as for the case $g=0$ \cite{curza}. 
 Note,  
however, that this argument is formal in the sense that it does not take 
into account the effects of renormalization.

\section{Renormalization of the dually related models}
\subsection{Coupling constant renormalization procedure}
To carry out explicitly the renormalization of the original and dual models,
Eq~.(\ref{lagrsig2}) and (\ref{dualag}), we use the general strategy 
developed in \cite{bfhp}. 
Since it is desribed there in quite some  
detail here we summarize only the main points.  
The procedure is based on the well known one and 
two loop counterterms \cite{hulto},\cite{metse},\cite{osb1} 
for the class of general bosonic $\sigma $ models, obtained by the background
field method in the dimensional regularization scheme. 
The  
counterterms of the Lagrangian
\be\label{slag}
{\cal L}={1\over2\lambda}\bigl( g_{ij}\pa_\mu\xi^i\pa^\mu\xi^j
+\epsilon_{\mu\nu}
b_{ij}\pa^\mu\xi^i\pa^\nu\xi^j\bigr)={1\over\lambda}\tilde{\cal L} ,
\ee
can be written as  
\be
\mu^{\epsilon}{\cal L}_1={\alpha^\prime\over2\epsilon\lambda}
\hat{R}_{ij}(\pa_\mu\xi^i\pa^\mu\xi^j
+\epsilon_{\mu\nu}\pa^\mu\xi^i\pa^\nu\xi^j)
={1\over\pi\epsilon}\Sigma_1 ,
\ee
and
\be
\mu^{\epsilon}{\cal L}_2={1\over2\epsilon}
({\alpha^\prime\over2})^2{1\over2\lambda}
Y^{lmk}_{\ \ \ j}\hat{R}_{iklm}(\pa_\mu\xi^i\pa^\mu\xi^j
+\epsilon_{\mu\nu}\pa^\mu\xi^i\pa^\nu\xi^j) 
={\lambda\over8\pi^2\epsilon}\Sigma_2,
\ee
where $\alpha^\prime=\lambda/(2\pi)$ 
\be\label{y}
\eqalign{
Y_{lmkj}=&-2\hat{R}_{lmkj}+3\hat{R}_{[klm]j}
+2(H^2_{kl}g_{mj}-H^2_{km}g_{lj})\,,\cr
H^2_{ij}=&H_{ikl}H_j^{kl}\,,\qquad
2H_{ijk}=\partial_ib_{jk}+{\rm cyclic}\,,
}\ee 
 and $\hat{R}_{ij}$
resp.\ $\hat{R}_{iklm}$ denote
the `generalized' (i.e. containing torsion) 
Ricci resp.\ Riemann tensors of the background $g_{ij}$ and $b_{ij}$. 
If the metric, 
$g_{ij}$, and the torsion 
potential, $b_{ij}$, depend also on a  parameter, $g$;
$g_{ij}=g_{ij}(\xi ,g)$, $b_{ij}=b_{ij}(\xi ,g)$, then the 
counterterms 
are converted into coupling, parameter
and (in general non-linear) field renormalizations:
\be\label{parren}
\eqalign{
\lambda_0&=\mu^\epsilon\lambda\Bigl( 1+{\zeta_1(g)\lambda\over\pi\epsilon}
+{\zeta_2(g)\lambda^2\over8\pi^2\epsilon}+...\Bigr)=\mu^\epsilon\lambda
Z_{\lambda}(g,\lambda),\cr   
g_0&= g+{x_1(g)\lambda\over\pi\epsilon}
+{x_2(g)\lambda^2\over8\pi^2\epsilon}+...=gZ_g(g,\lambda),\cr}
\ee
\be\label{fielren}
\xi^j_0=\xi^j+{\xi^j_1(\xi^k,g)\lambda\over\pi\epsilon}
+{\xi^j_2(\xi^k,g)\lambda^2\over8\pi^2\epsilon}+...,
\ee
if the $\zeta_i(g)$, $x_i(g)$ and $\xi_i^j(\xi^k,g)$ 
quantities solve the \lq\lq conversion equation":
\be\label{conv}
-\zeta_i(g)\tilde{\cal L}+{\pa \tilde{\cal L}\over\pa g}x_i(g)+
{\delta \tilde{\cal L}\over\delta\xi^k}\xi^k_i(\xi ,g)
+({\rm gauge})=\Sigma_i,\qquad i=1,2.
\ee
The appearance of (gauge) in Eqs.~(\ref{conv}) implies that the equality
of the two sides is required up to a (perturbative) gauge transformation of
the antisymmetric tensor field $b_{ij}$: 
\be
b_{ij}\rightarrow b_{ij}+({\rm gauge})=b_{ij}+
\frac{\lambda}{\pi\epsilon}\pa_{[i}W_{j]}^{(1)}(\xi^k,g)+
\frac{\lambda^2}{8\pi^2\epsilon}\pa_{[i}W_{j]}^{(2)}(\xi^k,g),
\ee
which changes $\tilde{\cal L}$, but leaves the torsion invariant. 
 ($[ij]=ij-ji$) 
We emphasize that it is not a priori guaranteed that Eqs.~(\ref{conv})
may be solved at all
for the unkown quantities.
If Eqs.~(\ref{conv}) do not have a solution, then
the renormalization of the model is not possible within the
restricted subspace
characterized by the coupling $\lambda$ and the parameter $g$, only   
in the (infinite dimensional)
space of all metrics and torsion potentials.
On the other hand, if Eqs.~(\ref{conv}) admit a solution, then the model is
renormalizable in the restricted field theoretic sense, and 
writing
$Z_\lambda=1+{y_\lambda (\lambda ,x)\over\epsilon}+...$ and 
$Z_g=1+{y_g(\lambda ,x)\over\epsilon}+...$, the $\beta$ functions of $\lambda$ 
and $g$, defined in the standard way,
$\beta_\lambda=\mu{d\lambda\over d\mu}$,
$\beta_g=\mu{dg\over d\mu}$,
 are readily obtained \cite{bfhp}:
\be\label{beta}
\beta_\lambda=\lambda^2{\pa y_\lambda\over\pa\lambda},\qquad 
\beta_g=g\lambda {\pa y_g\over\pa\lambda}.
\ee

In ref.\cite{bfhp}
it was shown that the deformed $SU(2)$ model, Eq.~(\ref{lagrsig2}), 
is
renormalizable in the ordinary sense both in the one and in two loop order:
 there is
no wave function renormalization for
$\theta$, $\phi$ and $\psi$, while the coupling constant
and the parameter get renormalized as in Eq.~(\ref{parren}); the solutions 
of Eq.~(\ref{conv}) finally lead to  
\be\label{betasig1}
\eqalign{
\beta
_\lambda=&-{\lambda^2 \over4\pi}\bigl( 1-g+{\lambda\over8\pi}(1-2g+5g^2)\bigr)\,
,\cr
\beta_g=&{\lambda\over2\pi}g(1+g)\bigl( 1+{\lambda\over4\pi}(1-g)\bigr)\,.
}\ee
Note that the $g=0$ resp.\ the $g=-1$ lines are
fixed lines under
the renormalization group, and $\beta_\lambda$
reduces to the $\beta$
function of the $SU(2)$ principal model, resp.\ of the $O(3)$
$\sigma$-model on them.
In the most interesting 
($\lambda\ge0$, $g<0$) quarter of the ($\lambda$,$g$) plane
the renorm trajectories
run into $\lambda=0$, $g=-1$; while for $g>0$ they run to infinity.

\subsection{One loop renormalization of the dual models}
To use Eqs.~(\ref{slag} -\ref{conv}) in the renormalization of the dual model, 
Eq.~(\ref{dualag}), we index the $\rho $, $\alpha $ and $z$ fields as 
$\xi^1$, $\xi^2$ and $\xi^3$. First we work in the one loop order only. 
The various components of the generalized 
Ricci tensor, that determine the one loop counterterm, $\Sigma_1$, 
are collected in the Appendix. Inspecting them we deduce the following: 

$\bullet $ as $\hat{R}_{ij}$ do not reproduce the 
polynomial form of the metric, 
$g_{ij}$, and torsion potential, 
$b_{ij}$,  of the dual model, Eq.~(\ref{dualag}), to abstract the 
coupling constant renormalization we have to assume that the    
$\rho $, $\alpha $ and $z$ fields undergo a (possibly nonlinear) 
renormalization like in Eq.~(\ref{fielren}) and also the gauge transformations
may be present. We denote the $\xi_1^k(\xi ,g)$ 
one loop corrections to $\rho $, $\alpha $ and $z$ as $F(\rho ,z,\alpha ,g)$,
$\alpha Y(\rho ,z,\alpha ,g)$ and 
$G(\rho ,z,\alpha ,g)$, respectively and also  
delete 
the index $1$ from $W_i(\rho ,z,\alpha ,g)$. 

$\bullet $ as $\hat{R}_{ij}$ do not depend on $\alpha $, (a manifestation, 
that 
the background field method preserves the symmetry translating $\alpha $), 
none of the $F$, $Y$ and $G$ functions may depend on $\alpha $.

$\bullet $ from the (anti)symmetry properties of $\hat{R}_{ij}$ it 
follows, that $\Sigma_1$ contains no new derivative couplings that are not 
present in ${\cal L}^d$, Eq.~(\ref{dualag}). In particular - as 
$\hat{R}_{12}=-\hat{R}_{21}$ and $\hat{R}_{32}=-\hat{R}_{23}$ - 
it contains no $\pa_\mu\rho\pa^\mu\alpha $ and $\pa_\mu z\pa^\mu\alpha $ 
terms. Therefore $Y(\rho ,z,g)$ may depend only on $g$, $Y=Y(g)$, as the only 
source of e.g. $\pa_\mu\rho\pa^\mu\alpha $ on the left hand side of 
Eq.~(\ref{conv}) is proportional to $\pa_\rho Y$. Furthermore, as 
$\hat{R}_{31}=\hat{R}_{13}$ we must have $\pa_1W_3-\pa_3W_1=0$; combining 
this with the $\alpha $ independence we put as an Ansatz $W_1=W_3=0$, 
$W_2=W(\rho ,z,g)$. 

Thus there are three functions of $\rho $, $z$ and $g$ (namely $F$, $G$
and $W$) and
three fuctions of $g$ ($\zeta_1(g)$, $x_1(g)$ and $Y(g)$) at our disposal to 
solve the algebro-differential system of equations originating from 
Eq.~(\ref{conv}). Introducing the notation
\be\label{ddef}
\pa_\rho F=FR,\quad \pa_z F=FZ,\quad \pa_\rho G=GR,\quad \pa_z G=GZ,
\ee
\be
\pa_1W_2-\pa_2W_1=\pa_\rho W=WR,\quad 
\pa_2W_3-\pa_3W_2=-\pa_z W=-WZ,
\ee 
and comparing the coefficients of $(\pa_\mu\rho )^2$, 
$\epsilon_{\mu\nu}\pa^\mu\rho\pa^\nu\alpha $, $(\pa_\mu\alpha )^2$, 
$\pa_\mu\rho\pa^\mu z$, \hfill\break 
$\epsilon_{\mu\nu}\pa^\mu\alpha\pa^\nu z $ and 
$(\pa_\mu z )^2$ on the two sides of Eq.~(\ref{conv}) one finds indeed:
\be\label{e11}
\eqalign{
&-{\frac {\zeta_1\left (1+\,g+\,\rho ^{2}\right )}{2\,D}}-{\frac {\left (1+\,g
+\,\rho ^{2}\right )\left (2\,\rho F+{\it x_1}\,\left (1+z^{2}\right )+2\,
\left (1+g\right )zG\right )}{2\,D^{2}}}\cr 
&+{\frac {2\,{\it x_1}\,+4\,\rho F+
4\,\left (1+g+\rho ^{2}\right ){\it FR}+4\,z\rho {\it GR}}{4\,D}}={\frac {
\hat{R}_{11}}{4}},}
\ee
\be\label{e12}
\eqalign{
&{\frac {\zeta_1z\rho \left (1+g\right )}{2\,D}}+{\frac {z\rho \left (1+g\right )
\left (2\,\rho F+{\it x_1}\,\left (1+z^{2}\right )+2\,\left (1+g\right )zG
\right )}{2\,D^{2}}}\cr 
&-{\frac {
(G\rho +zF+z\rho {\it FR}+z\rho {\it Y} )
\left (1+g\right )
+z\rho {\it x_1}\,-\rho ^{2}{\it GR}}{2\,D}}+\frac{WR}{2}
={\frac {\hat{R}_{12}}{4}},}
\ee
\be\label{e22}
\eqalign{
&-{\frac {\zeta_1\left (1+g\right )\rho ^{2}}{2\,D}}
-{\frac {\left (1+g\right )\rho ^
{2}\left (2\,\rho F+{\it x_1}\,\left (1+z^{2}\right )+2\,\left (1+g\right 
)zG\right )}{2\,D^{2}}}\cr 
&+{\frac {2\,{\it x_1}\,\rho ^{2}+4\,\left (1+g
\right )\rho F+4\,\left (1+g\right )\rho ^{2}{\it Y}}{4\,D}}
={\frac {\hat{R}_{22}}{4}},}
\ee
\be\label{e13}
\eqalign{
&-{\frac {\zeta_1z\rho }{2\,D}}-{\frac {z\rho \left (2\,\rho F+{\it x_1}\,\left (1+z^{2}
\right )+2\,\left (1+g\right )zG\right )}{2\,D^{2}}}+\cr 
&{\frac {\left 
(1+g+\rho ^{2}\right ){\it FZ}+\left (1+z^{2}\right ){\it GR}+G\rho +
zF+z\rho {\it FR}+z\rho {\it GZ}}{2\,D}}={\frac {\hat{R}_{13}}{4}},}
\ee
\be\label{e23}
\eqalign{
&{\frac {\zeta_1\rho ^{2}}{2\,D}}+{\frac {\rho ^{2}\left (2\,\rho F+{\it x_1}\,\left (1+z
^{2}\right )+2\,\left (1+g\right )zG\right )}{2\,D^{2}}}\cr 
&+{\frac {z\rho 
\left (1+g\right ){\it FZ}-2\,\rho F-\rho ^{2}{\it Y}-\rho ^{2}{\it GZ}}{2\,D}}
-\frac{WZ}{2}={\frac {\hat{R}_{23}}{4}},}
\ee
\be\label{e33}
\eqalign{
&-{\frac {\zeta_1\left (1+\,z^{2}\right )}{2\,D}}-{\frac {\left (1+\,z^{2}
\right )\left (2\,\rho F+{\it x_1}\,\left (1+z^{2}\right )+2\,\left (1+g
\right )zG\right )}{2\,D^{2}}}\cr 
&+{\frac {4\,zG+4\,\left (1+z^{2}\right )
{\it GZ}+4\,z\rho {\it FZ}}{4\,D}}={\frac {\hat{R}_{33}}{4}},}
\ee
(where $D=\rho^2+(1+g)(1+z^2)$). 
Note that this system contains $F$, $FR$, $FZ$, $G$, $GR$ and $GZ$ linearly, 
thus - apart from some pathological cases - these quantities may be determined 
from Eqs.(\ref{e11}-\ref{e33}) algebraically. 
Then we require that Eq.s~(\ref{ddef}) hold; i.e. 
that $FR$ be indeed $\pa_\rho F$, etc.. These requirements yield four 
equations that should determine $\zeta_1(g)$, $x_1(g)$ and $Y(g)$. Note 
however, that these four equations should be satisfied for all values of 
$\rho $ and $z$, thus it is not clear at all that a choice of 
$\zeta_1$, $x_1$ and $Y$, 
depending only on $g$, exists that guarantees this. This is the point 
where the gauge transformation described by $W_i$ plays an essential role,
as we must try to choose it in such a way that the four equations yield 
a $\rho $ and $z$ independent solution. 
We emphasize that the imposition of 
Eq.s~(\ref{ddef}) on the algebraic solution of 
Eq.s~(\ref{e11}-\ref{e33}) is the essential step 
of our renormalization program, as this step guarantees that the emerging new 
couplings can be accounted for by a nonlinear field redefinition. 

After some effort the procedure just described yields the following solution:
\be\label{ere1}
\eqalign{
F&={\rho(1-gz^2)\over2D}\,+\,{g-1\over4}\rho ,\quad Y=0,\cr
G&={z\big[(1+g)^2+g\rho^2\big]\over2D}\,+\,{g-1\over4}z,
\quad W=-{zg\rho^2\over4D}\cr}
\ee
\be\label{ere2}
\zeta_1(g)=-\frac{1-g}{4},\qquad {\it x}_1(g)=\frac{g(1+g)}{2}.
\ee
 Therefore, in the one loop order, 
${\cal L}^d$ may 
be renormalized in the restricted, field theoretic sense, i.e.  
 it 
is really possible to convert the counterterm $\Sigma_1$, 
into coupling constant 
and parameter renormalizations. Furthermore using 
the $\zeta_1$ and $x_1$ in Eq.s~(\ref{parren}) and (\ref{beta}), 
reproduces the one loop $\beta_\lambda$ of the original model in 
Eq.~(\ref{betasig1}).  
Thus as far as one 
loop coupling constant renormalization is concerned the equivalence between
the non Abelian dual and the original sigma model is established. 

Note that for $g=0$ the gauge contribution vanishes in 
Eq.s~(\ref{ere1},\ref{ere2}) while the $F$ and $G$ one loop 
corrections to $\rho $ and 
$z$ are the same as the ones obtained by the 
$\rho =r\cos\gamma$, $z=r\sin\gamma $ substitution from the renormalization 
of the \lq spherically symmetric' non Abelian dual of the $SU(2)$ principal 
model \cite{bfhp},\cite{subtu}.

Notice, however, that for $g=-1$, i.e. for the dual of the $O(3)$ model, the
gauge contribution does not vanish in spite of the decoupling of the 
$\alpha $ field in Eq.~(\ref{dualag}). Therefore it is not entirely clear, 
that starting with the simpler (purely metric) form of the dual Lagrangian, 
Eq.~(\ref{o3nad}), one would end up with the
solution in (\ref{ere1},\ref{ere2}): using ${\cal L}^d_{O(3)}$ 
and the $g\equiv -1$ form of
$\hat{R}_{ij}$ in eq.~(\ref{conv}) yields a system consisting of three 
equations only, instead of the six ones in (\ref{e11})-(\ref{e33}), and 
the gauge contribution is {\sl absent} as is the antisymmetric tensor field.  
To clarify 
this question and to confirm the results in (\ref{ere1},\ref{ere2}) 
we repeat the 
renormalization of the non Abelian dual of the $O(3)$ sigma model using 
${\cal L}^d_{O(3)}$, Eq.~(\ref{o3nad}). Since the emerging formulae become
much simpler this way, 
this also makes  possible to 
extend the analysis in this particular example to two loops.

\subsection{The dual of the $O(3)$ model at two loops}

Since the target space of this model is two dimensional, it is possible to 
find new variables, $x=x(\tau ,\sigma )$, $y=y(\tau ,\sigma )$ instead of 
$\rho $ and $z$,
\be\label{ujdef}
\rho =\Psi (x,y),\qquad z=\Gamma (x,y),
\ee
such that the target space metric is manifestly conformal to the flat one:
\be\label{flatm}
ds^2=d\rho^2+\frac{1+z^2}{\rho^2}dz^2+2\frac{z}{\rho }d\rho dz=
f^2(x,y)\bigl( dx^2+dy^2\bigr)\, .
\ee    
In terms of these fields, the Lagrangian, ${\cal L}^d_{O(3)}$, assumes the 
form
\be\label{lagrhar}
{\cal L}=\frac{1}{2\lambda}f^2(x,y)[(\pa_\mu x)^2+(\pa_\mu y)^2 ],
\ee
and the (ordinary) Ricci tensor, that determines $\Sigma_1$ in the absence of 
torsion, is readily obtained:
\be
\tilde{R}_{ij}=-\delta_{ij}(\pa_x^2+\pa_y^2)\ln f.
\ee
Notice, that $\tilde{R}_{ij}$, just like $g_{ij}$ is proportional to 
$\delta_{ij}$. Therefore there is no $\pa_\mu x\pa^\mu y$ term in $\Sigma_1$, 
and if we denote by $X=X(x,y)$ and $Y=Y(x,y)$ the $\xi_1^k(\xi )$ one loop
corrections (eq.~(\ref{fielren})) 
to $x$ and $y$ respectively, then the vanishing of 
$\pa_\mu x\pa^\mu y$ on the left hand side of Eq.~(\ref{conv}), using 
Eq.~(\ref{lagrhar}), requires:
\be\label{cr1}
\pa_y X=-\pa_x Y.
\ee
As $\tilde{R}_{ij}\sim\delta_{ij}$ the coefficients of $(\pa_\mu x)^2$ and 
$(\pa_\mu y)^2$ must be equal on the left hand side of Eq.~(\ref{conv}) too; 
from this it follows that:
\be\label{cr2}
\pa_x X=\pa_y Y.
\ee 
Eq.s~(\ref{cr1}) and (\ref{cr2}) imply that $X$ and $Y$ are harmonic functions, 
and are the real and imaginary parts of a holomorphic function $\epsilon (x,y)=
X+iY$. Finally, writing $f=\exp (\Phi)$, the only \lq non trivial' equation, 
following from Eq.~(\ref{conv}) and Eq.~(\ref{lagrhar}) is:
\be\label{renorm3}
-\frac{\zeta_1}{2}+\pa_x X+(X\pa_x\Phi +Y\pa_y\Phi )=
-\frac{1}{4}{\rm e}^{-2\Phi}(\pa_x^2+\pa_y^2)\Phi\, .
\ee
Thus the question of one loop renormalizability of ${\cal L}^d_{O(3)}$ can be 
formulated whether for the given $\Phi $ (see explicitely below) it is 
possible to choose the $\zeta_1$ constant such that (\ref{renorm3}) admits 
harmonic solutions $X$ and $Y$. 
  
To obtain the explicit form of $\Phi $ we have to find the mapping $\Psi 
(x,y)$ and $\Gamma (x,y)$, Eq.~(\ref{ujdef}). The second equality in 
Eq.~(\ref{flatm}) can be transformed into a system of differential equations
for 
\be
\frac{1}{2}\Psi^2(x,y)=K(x,y);\qquad\frac{1}{2}\Gamma^2(x,y)=H(x,y);
\ee
which admits the solution:
\be
H=a_0x^2,\qquad K=-a_0x^2+y\epsilon\sqrt{2a_0},
\ee
and leads finally to 
\be\label{fesfi}
f^2(x,y)=\frac{1}{-x^2+y\epsilon\sqrt{2/a_0}};\qquad 
\Phi =-\frac{1}{2}\ln (-x^2+y\epsilon\sqrt{2/a_0})\, .
\ee
(Here $a_0> 0$ is a constant of integration and the sign, $\epsilon =\pm $, 
is chosen so as to guarantee the positivity of $f^2$ in some domain. Note that
we use this mapping only locally, in an appropriate domain of $(x,y)$, the 
questions about the shape of this domain, its boundary etc. are beyond the scope
of this paper). Plugging this $\Phi $ into Eq.~(\ref{renorm3}) leads to  
\be\label{renorm4}
\bigl(-\frac{\zeta_1}{2}+\pa_x X\bigr)(-x^2+y\epsilon
\sqrt{\frac{2}{a_0}}) 
+Xx -\frac{\epsilon}{\sqrt{2a_0}}Y=
-\frac{1}{4}\bigl( x^2+y\epsilon
\sqrt{\frac{2}{a_0}}+\frac{1}{a_0}\bigr)\, .
\ee
This equation admits the solution:
\be
\zeta_1=-\frac{1}{2};\qquad X=-x;\qquad Y=-y+\frac{1}{2\epsilon\sqrt{2a_0}}\, .
\ee
Note that this $\zeta_1$ is the same as the one in Eq.~(\ref{ere2}) 
for $g=-1$. However, 
using the definition, $\rho =\Psi (x,y)=\sqrt{2K}$, 
$z=\Gamma (x,y)=\sqrt{2H}$, 
it is easy to show that the non linear redefinition described by this 
$X$ and $Y$ corresponds to an $F$ and $G$ different from the one 
in Eq.~(\ref{ere1}), reflecting the absence of torsion and the gauge
transformation. Since only the coupling constant and the parameter have 
physical significance we can state that  
the one loop renormalizability of the $O(3)$ model's non Abelian dual and 
Eq.~(\ref{ere2}) are  confirmed.

The next logical step is to investigate the renormalizability of this model 
in the two loop order. (The two loop non renomalizability of the non Abelian 
dual of the $SU(2)$ principal model -- which is the only other case of the
deformed sigma models when the complexity of this problem becomes tractable 
-- is discussed in \cite{bfhp},\cite{subtu}). 
Computing 
$Y^{lmk}_{\ \ \ j}\tilde{R}_{iklm}$ for the metric $f^2(x,y)[(dx)^2+(dy)^2]$ 
one finds it to be proportional to $\delta_{ij}$, thus $\Sigma_2$ containes 
no $\pa_\mu x\pa^\mu y$ term either, and Eq.s~(\ref{cr1} ,\ref{cr2}) are also 
valid for the $\tilde{X}(x,y)$, $\tilde{Y}(x,y)$ two loop corrections to 
$x$ and $y$. Using the explicit form of $\Sigma_2$, the two loop equation 
following from Eq.~(\ref{conv}) can be written as: 
\be\label{twoloop}
\bigl(-\frac{\zeta_2}{2}+\pa_x \tilde{X}\bigr)(-x^2+y a)^2 
+(-x^2+y a)(x\tilde{X} -\frac{a}{2}\tilde{Y})=
\frac{1}{8}\bigl( a^2 +2x^2+2ya\bigr)^2\, ,
\ee
(where $a=\epsilon\sqrt{2/a_0}$). 
Taking into account the polynomial nature of this equation, (i.e. the fact 
that the coefficients of $\tilde{X}$, $\tilde{Y}$, $\pa_x\tilde{X}$, and 
also the terms independent of $\tilde{X}$ and $\tilde{Y}$ are finite, well 
defined polynomials in $x$ and $y$), it is not difficult to see that there is 
no choice of $\zeta_2$ that would make possible to find a pair of harmonic 
$\tilde{X}$ and $\tilde{Y}$ solving Eq.~(\ref{twoloop}).  Notice e.g. that 
the absence of terms containig $x^l$, $l >4$ and $y^m$, $m > 2$ among the 
 $\tilde{X}$, $\tilde{Y}$ independent terms in Eq.~(\ref{twoloop}) makes the 
coefficient of all the $k>1$ terms vanish in the natural polynomial Ansatz:
$\tilde{\epsilon}(x,y)=\tilde{X}+i\tilde{Y}=\sum\limits_{k=0}^{N}b_kw^k$;
 $w=x+iy$. The possibility 
  of $\tilde{\epsilon}$ being linear in $x$ and $y$ is 
eliminated by realizing that the matching of the various $x^ly^m$ terms on 
the two sides of (\ref{twoloop}) leads to mutually inconsistent expressions 
for $\zeta_2$. (The case of a rational $\tilde{\epsilon}$, 
$\tilde{\epsilon}=\frac{P_N(w)}{Q_M(w)}$, can be ruled out by a similar 
argument).       
Therefore we conclude that   
${\cal L}^d_{O(3)}$ is not renormalizable in the two loop order in the 
restricted, field theoretic sense.

\section{Discussion and conclusions}
In this paper we considered a one parameter family of sigma models 
interpolating between the $SU(2)$ principal model and the $O(3)$ sigma 
model together with the non Abelian dual of this family, and investigated the 
renormalization of the coupling constant and the deformation parameter in the 
two families of models. The interest of the $O(3)$ sigma model and its non 
Abelian dual stems from the fact that the fields of the former parametrize 
a coset space while usually non Abelian duality is formulated for sigma 
models defined on group manifolds; see however ref.\cite{klim3} for Poisson 
Lie duality in case of cosets.

Classically these two sets of models are related by
a canonical transformation, thus they are equivalent. If this equivalence 
persisted for the quantized models then the coupling constant and the parameter
of the original and dual models should be renormalized in the same way - 
apart from some potential change in the renormalization scheme. 

We found that in the {\sl one loop} order of perturbation 
theory this expected equivalence shows up for the complete family of models.
However, for the two particular models 
(for $g=0$ and $g=-1$, that describe the $SU(2)$ 
principal model and the $O(3)$ sigma model respectively), when the two loop 
analysis becomes tractable 
we found the equivalence broken in this order. We came
to these conclusions by establishing that the system of equations, 
guaranteeing 
that the coupling and parameter renormalizations can be extracted in the 
ususal, field theoretic sense from the counterterms of the sigma models, 
 have 
such a solution for the {\sl dual} model in the one loop order 
that leads to the same $\beta $ 
functions as the original model. However, in the two loop order
for $g=0$ or $-1$ at least, these equations have no solutions for the duals, 
thus the duals are not even renormalizable, hence the equivalence is 
obviously broken. We emphasize that the essential point is the non 
equivalence of the dually related models at the two loop level and we are 
using renormalizability only as a tool to show this.    

One may think that the reason behind the two loop discrepancy 
between the original 
and dual theories is the same as in the case of Abelian duality \cite{bfhp2}, 
namely, 
that the bare and renormalized quantities do not transfrom in the same way 
under duality transformations. In this respect 
the fact that the dual Lagrangian, ${\cal L}^d$, is equivalent to the 
original one for the complete family of deformed sigma models gives support 
to the idea that as far as coupling constant renormalization is considered 
non Abelian duality is similar to Abelian duality;  
as shown in ref.s \cite{bfhp},\cite{haag}, 
\cite{bfhp3}, for models connected by these transformations, 
in the one loop order, (after carrying out the required, usually 
highly non trivial field renormalizations in the duals), the coupling
constants and the parameters are renormalized in the same way. If, for 
non Abelian duality, this indeed turns out to be true in general, then one 
can go on and look for the required modifications of the transformation 
rules for the {\sl renormalized} quantities in the framework outlined in 
\cite{bfhp2}.

\noindent{\bf Acknowledgements} L.P. thanks C. Klimcik for 
useful discussions and a correspondence. J.B. thanks for the Bourse de la 
Region Centre, No 96 298 027. 
This investigation was supported 
in part by the Hungarian National Science Fund (OTKA) under 
T016251 and T019917.
  
\section{Appendix} 
Here we collect the components of the generalized Ricci tensor of the 
dual model, Eq.~(\ref{dualag}). 
\be
\eqalign{
\hat{R}_{11}=&
-\frac {1}{2\,
\left (1+z^{2}+g+gz^{2}+\rho ^{2}\right )^{3}}\Bigl(
-3+2\,gz^{2}\rho ^{4}+8\,\rho ^{2}z^{2}g^{3}\cr &+3\,g^{2}z^{4}\rho ^{2}+g^{3}
z^{4}\rho ^{2}+3\,z^{4}g\rho ^{2}-6\,g+12\,gz^{2}+6\,\rho ^{2}z^{2}+
3\,g\rho ^{2}+2\,z^{4}g\cr &+28\,g^{2}z^{2}+8\,g^{2}z^{4}-3\,g^{2}\rho ^{2}+
20\,g^{3}z^{2}+6\,g^{3}z^{4}+20\,gz^{2}\rho ^{2}+22\,g^{2}z^{2}\rho ^{2}
\cr &-z^{4}-4\,g^{2}+7\,\rho ^{4}-2\,g^{3}+3\,\rho ^{2}+4\,\rho ^{4}z^{2}g^{2}+
\rho ^{2}z^{4}\cr &+2\,\rho ^{4}z^{2}
-3\,g^{3}\rho ^{2}-3\,g^{2}\rho ^{4}+4\,z^{2}g^{4}-g\rho ^{6}+g^{4}z^{4}-g^{4}
+\rho ^{6}\Bigr)}
\ee
\be
\hat{R}_{12}=-\hat{R}_{21}=
-{\frac {2\,\rho \left (1+g\right )z\left ((1+g)^{3}+2\,g^{2}\rho ^{2}+gz
^{2}\rho ^{2}+3\,g\rho ^{2}+g\rho ^{4}
\right )}{\left (1+z^{2}+g+gz^{2}+\rho ^{2}\right )^{3}}}
\ee
\be
\eqalign{
\hat{R}_{22}=
&-\frac {\left (1+g\right )^{2}\rho ^{2}}
{2\,\left (1+z^{2}+g+gz^{2}+\rho ^{2}\right )^{3}}
\bigl(-g^{2}+4\,g^{2}z^{2}+g^{2}z^{4}+
4\,z^{4}g+12\,gz^{2} \cr &-2\,g\rho ^{2}+4\,gz^{2}\rho ^{2}-
3-z^{4}-2\,\rho ^{2}z^{2}-\rho ^{4}\bigr) }
\ee
\be
\eqalign{
\hat{R}_{13}=&\hat{R}_{31}=
-\frac{z\rho }{2\,\left (1+z^{2}+g+gz^{2}+\rho ^{2}\right )^{3}}
\bigl( 5\,g+8\,\rho ^{2}+8\,z^{2}\cr &
+20\,gz^{2}+2\,\rho ^{2}z^{2}+g^{2
}+2\,g\rho ^{2}+3\,z^{4}g+16\,g^{2}z^{2}+3\,g^{2}z^{4}+2\,gz^{2}\rho ^{2}\cr 
&+z^{4}+\rho ^{4}-g^{3}-2\,g^{2}\rho ^{2}+4\,g^{2}z^{2}\rho ^{2}
+4\,g^{3}z^{2}+g^{3}z^{4}-g\rho ^{4}+3\bigr) }
\ee
\be
\hat{R}_{23}=-\hat{R}_{32}=
\frac{2\left (1+g\right )\rho ^{2}}{\left (1+z^{2}+g+gz^{2}+\rho ^{2}\right )
^{3}}
\left ( g^{2}z^{2}+z^{4}g+gz^{2}\rho ^{2}+3\,gz^{2}-1
\right )
\ee
\be
\eqalign{
\hat{R}_{33}=&
-\frac{1}{2\,\left (1+z^{2}+g+gz^{2}+\rho ^{2}\right )^{3}}\bigl( 
-3-gz^{2}\rho ^{4}+4\,g^{2}z^{4}\rho ^{2}+2\,z^{4}g\rho ^{2}\cr 
&-9\,g+9\,gz^{2}+6\,\rho ^{2}z^{2}
-10\,g\rho ^{2}+21\,z^{4}g+9\,g^{2}z^{2}+21\,g^{2}z^{4}-6\,g^{2}\rho ^{2}\cr 
&+3\,g^{3}z^{2}+7\,g^{3}z^{4}-3\,g\rho ^{4}
+2\,g^{2}z^{2}\rho ^{2}+
7\,z^{4}-9\,g^{2}-\rho ^{4}-3\,g^{3}\cr 
&+3\,z^{2}+3\,z^{6}g+3\,z^{6}g^{2}+g^{3
}z^{6}+z^{6}+2\,\rho ^{2}z^{4}+\rho ^{4}z^{2}\bigr) }
\ee
For $g=0$ these expressions simplify and become identical to the well known  
expressions \cite{bfhp},\cite{subtu},\cite{curza2} 
for the non Abelian dual of the $SU(2)$ principal model. For 
$g=-1$ $\hat{R}_{12}$, $\hat{R}_{22}$, and $\hat{R}_{23}$ vanish identically; 
this is in accord with the decoupling of the $\alpha $ field.

\end{document}